\documentclass[conference]{IEEEtran}
\usepackage{cite}
\usepackage{amsbsy}
\usepackage{url}
\usepackage{amsmath,amsthm,graphicx}
\usepackage{subfigure}
\usepackage{algpseudocode}

\usepackage{multirow}

\begin{document}
%
\title{Visualisation of Survey Responses using Self-Organising Maps:  A Case Study on Diabetes Self-care Factors}
\author{
    \IEEEauthorblockN{Santosh Tirunagari\IEEEauthorrefmark{1}, Simon Bull\IEEEauthorrefmark{1}, Samaneh Kouchaki \IEEEauthorrefmark{2}, Deborah Cooke \IEEEauthorrefmark{3} and Norman Poh \IEEEauthorrefmark{1}}
    \IEEEauthorblockA{\IEEEauthorrefmark{1} Department of Computer Science.}
    \IEEEauthorblockA{\IEEEauthorrefmark{3} School of Health Sciences.}
\IEEEauthorblockA{\IEEEauthorrefmark{1}\IEEEauthorrefmark{3}University of Surrey, Guildford, Surrey, United Kingdom GU2 7XH.}
\IEEEauthorblockA{\IEEEauthorrefmark{2} Evolutionary and Genome Sciences, Faculty of Biology, Medicine and Health, The University of Manchester.}
    \\\{s.tirunagari, s.c.bull, d.cooke, n.poh\}@surrey.ac.uk and samaneh.kouchaki@manchester.ac.uk.

}


\maketitle

\begin{abstract}
Due to the chronic nature of diabetes, patient self-care factors play an important role in any treatment plan. In order to understand the behaviour of patients in response to medical advice on self-care, clinicians often conduct cross-sectional surveys. When analysing the survey data, statistical machine learning methods can potentially provide additional insight into the data either through deeper understanding of the patterns present or making information available to clinicians in an intuitive manner. In this study, we use self-organising maps (SOMs) to visualise the responses of patients who share similar responses to survey questions, with the goal of helping clinicians understand how patients are managing their treatment and where action should be taken. The principle behavioural patterns revealed through this are that: patients who take the correct dose of insulin also tend to take their injections at the correct time, patients who eat on time also tend to correctly manage their food portions and patients who check their blood glucose with a monitor also tend to adjust their insulin dosage and carry snacks to counter low blood glucose. The identification of these positive behavioural patterns can also help to inform treatment by exploiting their negative corollaries.
\end{abstract}
\IEEEpeerreviewmaketitle
\section{Introduction}

The incidence of diabetes is a growing health problem worldwide. Caused by the presence of excess glucose within the blood, diabetes comes in two main types: type 1, where destruction of pancreatic beta cells leaves the body unable to produce enough insulin, and the more prevalent type 2, where the body is incapable of producing enough insulin due to insulin resistance. Complications concomitant with diabetes include retinopathy, nephropathy, neuropathy, stroke and myocardial infarction amongst others~\cite{NathanLongTermDMComps}. In the case of type 1 diabetics, it is necessary for patients to carefully monitor and manage their own insulin levels by monitoring their food intake and timing of meals, usually with accompanying insulin therapy administered via subcutaneous injections or an insulin pump. Due to the importance of these self-care factors, understanding how and why patients follow medical advice concerning diabetes self-care can improve the ability of clinicians to manage their patients. In this study, we consider the results of a survey collected to assess the self-care behaviours of diabetic patients, and attempt to identify and understand patients with similar patterns of self-care behaviour.

Health care survey data contains a wealth of information about patient behaviours. As it is usually heterogeneous and contains missing values, applying traditional statistical methods may give clinicians an oversimplified view of the data. Therefore, statistical machine learning tools, such as self organising maps (SOMs), can be utilised for performing data analysis. In the case of SOMs, clustered responses in the higher dimensional input space are mapped into a 2-dimensional grid for visualisation. SOMs can therefore be used to visually explore datasets and mine correlations within them.

\begin{table*}[]
\centering
\caption{Survey questions on Self-care factors}
\label{labels-self-factors}
\begin{tabular}{|l|l|l|}
\hline
Label &                    Questions based on                &  Self-care factors (Questions)                                                            \\ \hline
CBG-Monitor      & \multirow{3}{*}{Blood glucose}     &  Check blood glucose with monitor                                                         \\ \cline{1-1} \cline{3-3} 
RBG-Results      &                                    &  Record blood glucose results                                                             \\ \cline{1-1} \cline{3-3} 
CKGL-High        &                                    &  Check ketones when glucose level is high                                                 \\ \hline
TCD-Insulin      & \multirow{3}{*}{Insulin}           &  Take correct dose of insulin                                                             \\ \cline{1-1} \cline{3-3} 
TI-Time          &                                    &  Take insulin at the right time                                                           \\ \cline{1-1} \cline{3-3} 
AIDGFE           &                                    &  Adjust insulin dosage based on glucose values, food, and exercise                       \\ \hline
EC-Food Portions & \multirow{7}{*}{Lifestyle}         &  Eat the correct food portions                                                            \\ \cline{1-1} \cline{3-3} 
Eat-Timely       &                                    &  Eat meals/snacks on time                                                                 \\ \cline{1-1} \cline{3-3} 
Rec-Carbs        &                                    &  Treat low blood glucose with just the recommended amount of carbohydrate                  \\ \cline{1-1} \cline{3-3} 
Carry-Sugar      &                                    &  Carry quick acting sugar to treat low blood glucose                                      \\ \cline{1-1} \cline{3-3} 
KF-Records       &                                    &  Keep food records                                                                        \\ \cline{1-1} \cline{3-3} 
RF-Labels        &                                    &  Read food labels                                                                         \\ \cline{1-1} \cline{3-3} 
Exercise         &                                    &  Exercise                                                                                 \\ \hline
Clin-Appoint     & \multirow{2}{*}{Administrative}    &  Come in for clinic appointments                                                          \\ \cline{1-1} \cline{3-3} 
WM-Alert         &                                    &  Wear a medic alert ID                                                                    \\ \hline
\end{tabular}
\end{table*}

Previously, SOMs have been used to visually explore financial~\cite{deboeck1998visual}, gene expression~\cite{tamayo1999interpreting}, linguistics~\cite{honkela1997websom} and medical survey~\cite{tirunagari2014patient} datasets. Toronen $et$ $al.$~\cite{toronen1999analysis} used SOMs to analyse yeast gene expression data, and demonstrated its suitability for analysing and visualising gene expression profiles. Mehmood $et$ $al$.~\cite{mehmood2011self} found correlations between countries and nutrition, health and lifestyle using SOM analysis, while Tirunagari $et$ $al.$~\cite{tirunagari2012mining} used SOM analysis to correlate causal factors associated with different types of maritime accidents. It has also been shown that SOMs can be used to identify co-morbidities, link self-care factors that are dependent on each other and to visualise individual patient profiles in clinically useful ways, e.g. by giving a visual summary of individual patient profiles and enabling patients to be grouped together~\cite{tirunagari2014patient}. SOMs are therefore a potentially valuable tool for visually analysing high-dimensional medical questionnaire responses, as well as a means to visually summarise an individual patient's data.

The main objectives of this study are two-fold. Firstly, from the computational perspective, to examine the feasibility of using SOMs as a means of extracting  useful information from medial surveys. Secondly, from the scientific perspective, we would like to understand if the responses collected from the survey are reasonable and correspond to what clinicians would expect, for example by determining how the behaviours of patients group them together and how self-care factors are correlated.

Our contributions can thus be summarised as follows:
\begin{itemize}
\item {\bf Visualisation of diabetic self-care patient behaviours from medical survey data.}~ Although SOMs have been widely used in exploring survey data, their use in visualising  similar patient profiles is rarely discussed. We therefore explore the capabilities of SOMs to visualise patient profiles from survey data in a manner that is easy to interpret.
\item {\bf Clinical understanding.}~ We will explore the ability of visual analytics provided by SOMs to enhance experts' understanding of the impact of self-care behaviours in patients. 
\end{itemize}

The organisation of the paper is as follows: In section~\ref{data}, we present the survey dataset. In section~\ref{methods}, we present the SOM methodology. Experiments and results are discussed in section~\ref{expres}, including data preprocessing and imputation. Finally, in section~\ref{condis}, we discuss the conclusions that can be drawn from the results.

\section{Dataset}
\label{data}

The survey consists of 611 patients' responses to 15 questions on self-care factors across four categories as shown in Table~\ref{labels-self-factors}. \textbf{Blood glucose:} Regular monitoring of blood glucose levels is essential in any treatment plan for type 1 diabetics. Without this the patient will be unaware of any sudden spikes or drops in blood glucose, or how well their treatment plan is succeeding in managing their glucose levels. \textbf{Insulin:} Insulin is necessary for the body to effectively remove glucose from the blood. As type 1 diabetics produce insufficient insulin, they must usually supplement their lifestyle-based management of the disease with injections or an insulin pump. For those patients injecting insulin, taking the correct dose at the correct time, and modifying the dose appropriately, is necessary to ensure their blood glucose is managed adequately. \textbf{Lifestyle:} Careful control of diet is one aspect of lifestyle-based management of diabetes. For a patient, this involves not only eating the correct foods in the correct amounts, but also ensuring that a regimented schedule of meal times is followed accurately. Physical activity is the second aspect involved in managing diabetes via lifestyle choices. Exercise is beneficial for maintaining a healthy weight, thereby reducing the chance of complications, and helps control blood glucose levels by causing glucose to leave the blood due to the increased rate of respiration. \textbf{Administrative:} Attending clinic appointments is important for evaluating and updating a patient's treatment plan, and heading off the development of complications. Items like medic alert bracelets, while not used to control blood glucose, ensures that medics will be aware of a patient's diabetes, and is another measure of how conscientious they are in following their treatment plan.

The responses for the self-care behaviours are required to be one of the following:  1) Never, 2) Rarely, 3) Sometimes, 4) Usually and 5) Always.

\subsection{Data Preprocessing}
The processing of the data included the conversion of free text (yes, no, always, usually, sometimes, rarely, never, etc.), missing values (flagged as Not-a-Number, NaN) and Not Applicable (NA) answers into categorical numerical values. NA answers were scored as a 0, while free text values were mapped to one of the five answer categories and then scored such that always, often, sometimes, rarely and never were set to 5, 4, 3, 2 and 1 respectively. K-nearest neighbour based imputation (via Matlab's `\emph{knnimpute(Data)}') was used to replace NaN answers with an appropriate value. This was done by finding the patient nearest to the one with the missing value (using Euclidean distance), and replacing the missing value with that of the nearest neighbour. If the corresponding value from the nearest neighbour was also NaN, then the second nearest neighbour was considered.

\begin{table}[h]
\centering
\caption{Ethnic, martial and employment status of respondents.}
\label{tab:veracity}
\begin{tabular}{lll}
\hline
\textbf{Ethnicity} & \textbf{Count} & \textbf{Percent} \\ \hline
White British      & 558            & 91.33\%          \\
Other White        & 32             & 5.24\%           \\
Mixed              & 2              & 0.33\%           \\
British Asian      & 3              & 0.49\%           \\
Black British      & 4              & 0.65\%           \\
NA                 & 4              & 0.65\%           \\
Other Ethnic Group & 3              & 0.49\%           \\
Asian              & 3              & 0.49\%           \\
Black              & 2              & 0.33\%           \\ 
\\\hline
\textbf{Marital}              & \textbf{Count} & \textbf{Percent} \\ \hline
In a significant relationship & 467            & 76.43\%          \\
Single                        & 95             & 15.55\%          \\
Divorced / Separated          & 36             & 5.89\%           \\
NA                            & 5              & 0.82\%           \\
Widowed                       & 8              & 1.31\%       \\   
\\\hline
\textbf{Employer}                           & \textbf{Count} & \textbf{Percent} \\ \hline
Self-employed / Freelance without employees & 57             & 9.33\%           \\
Employee                                    & 515            & 84.29\%          \\
Self-employed with employees                & 20             & 3.27\%           \\
NA                                          & 19             & 3.11\%          
\end{tabular}
\end{table}

\subsection{Data Demographics}

\begin{figure}[ht]
\centering
\includegraphics[width=2.2in]{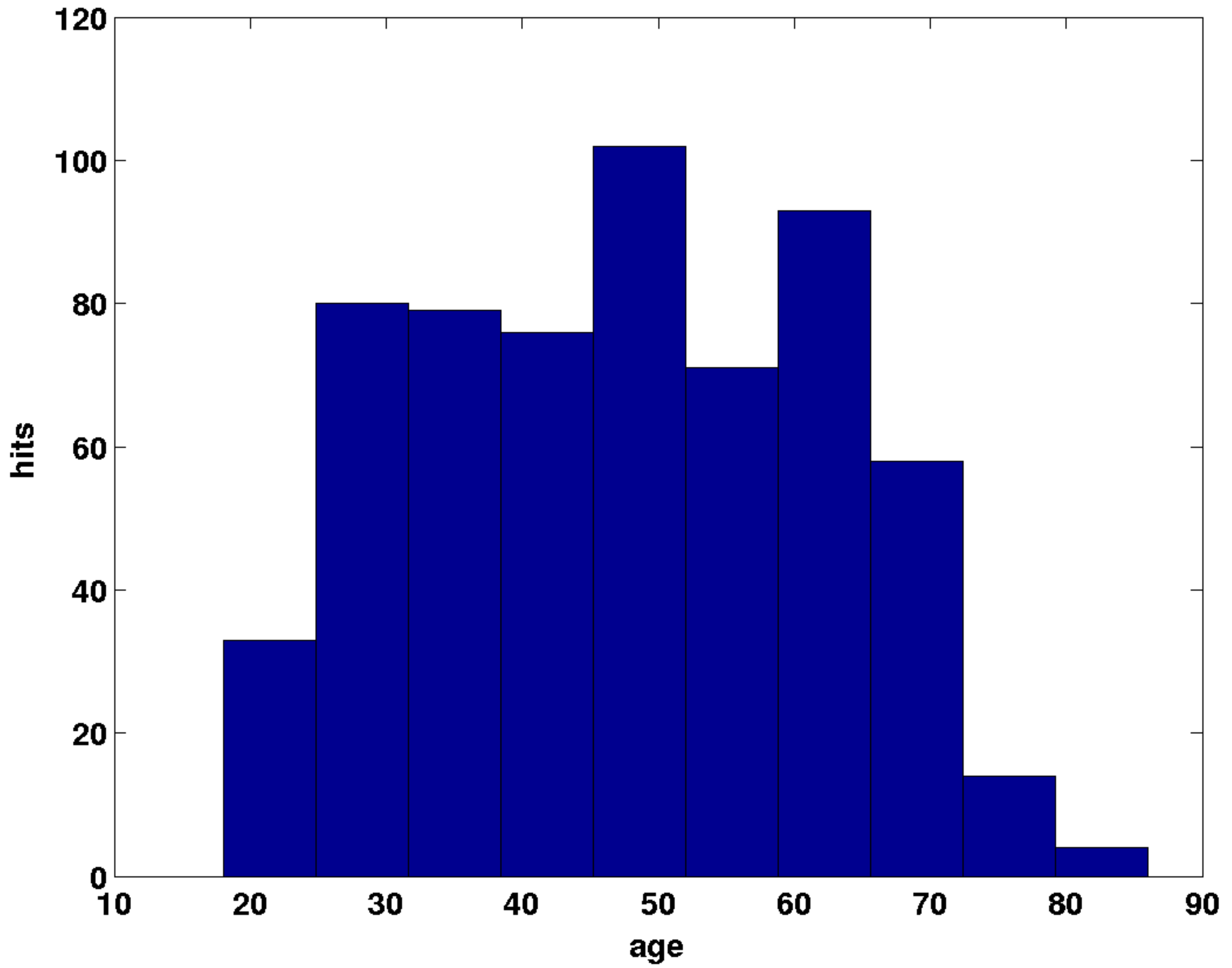}
\caption{Distribution of respondent ages.}
\label{fig_age}
\end{figure}

Participants in the survey were required to be at least eighteen years of age, with the eldest participants being over eighty. Respondents ages are evenly distributed from the mid-twenties through to approximately seventy years of age, with few patients outside this range (Figure~\ref{fig_age}).

From Table~\ref{tab:veracity}, we can see that the survey respondents are highly imbalanced in terms of their ethnicity, marital and employment status. Particularly in the case of ethnicity, where British respondents constitute 91.33\% of the cohort. Males constituted 45\% of the survey population and females 55\%. The most common employment status was full time employment (48.45\%) and the least unemployed (3.11\%). Modern professional occupations (23.4\%) and clerical and intermediate occupations (19.64\%) were the most common professions (Table ~\ref{demographics}).

\begin{table}[h]
\centering
\caption{Educational attainment, employment status and profession of respondents.}
\label{demographics}
\begin{tabular}{lll}
\hline
\textbf{Education level}                           & \textbf{Count} & \textbf{Percent} \\ \hline
Higher National Certificate (HNC) or Diploma (HND) & 62             & 10.15\%          \\
Undergraduate degree                               & 119            & 19.48\%          \\
'A' Levels or 'AS' Levels                          & 66             & 10.80\%          \\
Postgraduate degree                                & 107            & 17.51\%          \\
CSE, GCSE or 'O' Level                             & 138            & 22.59\%          \\
Other                                              & 63             & 10.31\%          \\
NA                                                 & 1              & 0.16\%           \\
No qualifications                                  & 55             & 9.00\%          \\
\\\hline
\textbf{Employment}        & \textbf{Count} & \textbf{Percent} \\ \hline
In part time employment    & 85             & 13.91\%          \\
In full time employment    & 296            & 48.45\%          \\
Unemployed                 & 19             & 3.11\%           \\
Looking after house/family & 27             & 4.42\%           \\
Retired                    & 128            & 20.95\%          \\
Student                    & 20             & 3.27\%           \\
Permanently sick/disabled  & 33             & 5.40\%           \\
NA                         & 3              & 0.49\%          \\
\\\hline
\textbf{Profession}                         & \textbf{Count} & \textbf{Percent} \\ \hline
Technical and craft occupations             & 46             & 7.53\%           \\
Modern professional occupation              & 143            & 23.40\%          \\
Traditional or professional occupation      & 69             & 11.29\%          \\
Senior managers or administrators           & 72             & 11.78\%          \\
Middle or junior manager                    & 40             & 6.55\%           \\
Other                                       & 10             & 1.64\%           \\
Clerical and intermediate occupations       & 120            & 19.64\%          \\
Routine manual and service occupations      & 27             & 4.42\%           \\
Semi routine manual and service occupations & 57             & 9.33\%           \\
NA                                          & 27             & 4.42\%          
\end{tabular}
\end{table}

The majority of the patients surveyed responded that the symptoms of hypoglycemia occurred at blood glucose levels of greater than or equal to 3.0mmol/L (Table~\ref{hypoglycemia}). The percentage of respondents that are  aware of hypoglycemia commencing was over half, with 51.88\% of patients being aware. 

\begin{table}[h]
\centering
\caption{Breakdown of patient responses to hypoglycemia questions.}
\label{hypoglycemia}
\begin{tabular}{lll}
\hline
\textbf{Hypoglycemia levels}               & \textbf{Count} & \textbf{Percent} \\ \hline
Less than 3.0 mmol/L               & 267            & 43.70\%          \\
Greater than or equal to 3.0mmol/L & 317            & 51.88\%          \\
Do not feel symptoms               & 22             & 3.60\%           \\
NA                                 & 5              & 0.82\%           \\
\\\hline
\textbf{Hypoglycemia awareness} & \textbf{Count} & \textbf{Percent} \\ \hline
Unaware   & 289   & 47.30\% \\
Aware     & 317   & 51.88\% \\
NA             & 5     & 0.82\% 
\end{tabular}
\end{table}

\section{Self-Organising Maps}
\label{methods}
A SOM, also known as Kohonen map, is an unsupervised artificial neural network technique used to produce a low dimensional representation of a high dimensional input space. This process of reducing the dimensionality of the data is based on a data compression technique known as vector quantisation. In addition, the high to low dimensional mapping is found such that any topological relationships within the high dimensional space are preserved after mapping to the low dimensional one. SOMs are therefore useful for visualising large high dimensional datasets. Rather than using error-correction in the manner of most neural networks (e.g. back-propagation with gradient descent), SOMs employ a competitive learning approach in which only one neuron wins at each training phase. Additionally, there are no intra-layer connections in either the input or output layers, however the neurons in the output layer do communicate with each other via a neighbourhood function, with the winning neuron in a training phase influencing its neighbours.

Consider the input vector $\boldsymbol{x}$
\begin{equation}
\boldsymbol{x}=[x_1,x_2, ..., x_n]^T
\end{equation}
The synaptic weight vector of the neuron $i$ in the 2-dimensional output layer of neurons is
\begin{equation}
\boldsymbol{w_i}=[w_i^1, w_i^2, ..., w_i^m]^T, i=1,2, \ldots, m
\end{equation}
where $m$ is the number of output neurons and $w_i^v$ is the weight associated with neuron $i$ and variable $v$. Although, the output neurons are arranged in a $2$-dimensional array, their weight vectors are $n$-dimensional, i.e. they have the same number of dimensions as the input vector $\boldsymbol{x}$. The Euclidean distance $\|\boldsymbol{x}-\boldsymbol{w_i}\|_2$ of the current input vector $\boldsymbol{x}$ to all the weight vectors $\boldsymbol{w_i}, i=1,2, \ldots, m$ is computed. The winning neuron is one whose weight vector $\boldsymbol{w_q}$ has the minimum Euclidean distance to $\boldsymbol{x}$, i.e.,
\begin{equation}
q(x)=arg_{min_i} \|\boldsymbol{x}-\boldsymbol{w_i}\|_2
\end{equation}
The weight vectors of the winning neuron and the neurons in its predefined neighbourhood
$\eta_q$ are updated using gradient descent, leading to the following update rule:
\begin{equation}
 w_i (k + 1) = w_i (k) + \eta_{qi}(k)[x(k) - w_i (k)]
\end{equation}
Neurons outside the neighbourhood are not updated, i.e $\eta_q(k)=0$, and neurons inside the neighbourhood  $\eta_q$ are updated using equation (\ref{eq1}):
\begin{equation}
\label{eq1}
\eta_{qi}(k)=\mu(k).
\end{equation} 
The learning parameter $\mu(k)$, with $0< \mu (k)<1$, is inversely proportional to the number of training iterations performed, and therefore decreases as the number of iterations increases. The learning process consists of an ordering, or rough training, phase and a convergence, or fine tuning, phase. In the ordering phase, the topological ordering of the weight vectors is performed. The learning parameter $\mu (k) $ is set close to unity. In the convergence phase, the self-organising map is fine-tuned. This is achieved by setting the learning parameters $ \mu (k)$  on the order of 0.01. The stopping criterion for the SOM algorithm is based on a specified number of iterations or minimum level of change in the weight vectors.

\section{Experiments and Results}
\label{expres}
\begin{figure}[h]
\centering
\begin{tabular}{c}
\includegraphics[width=3.5in]{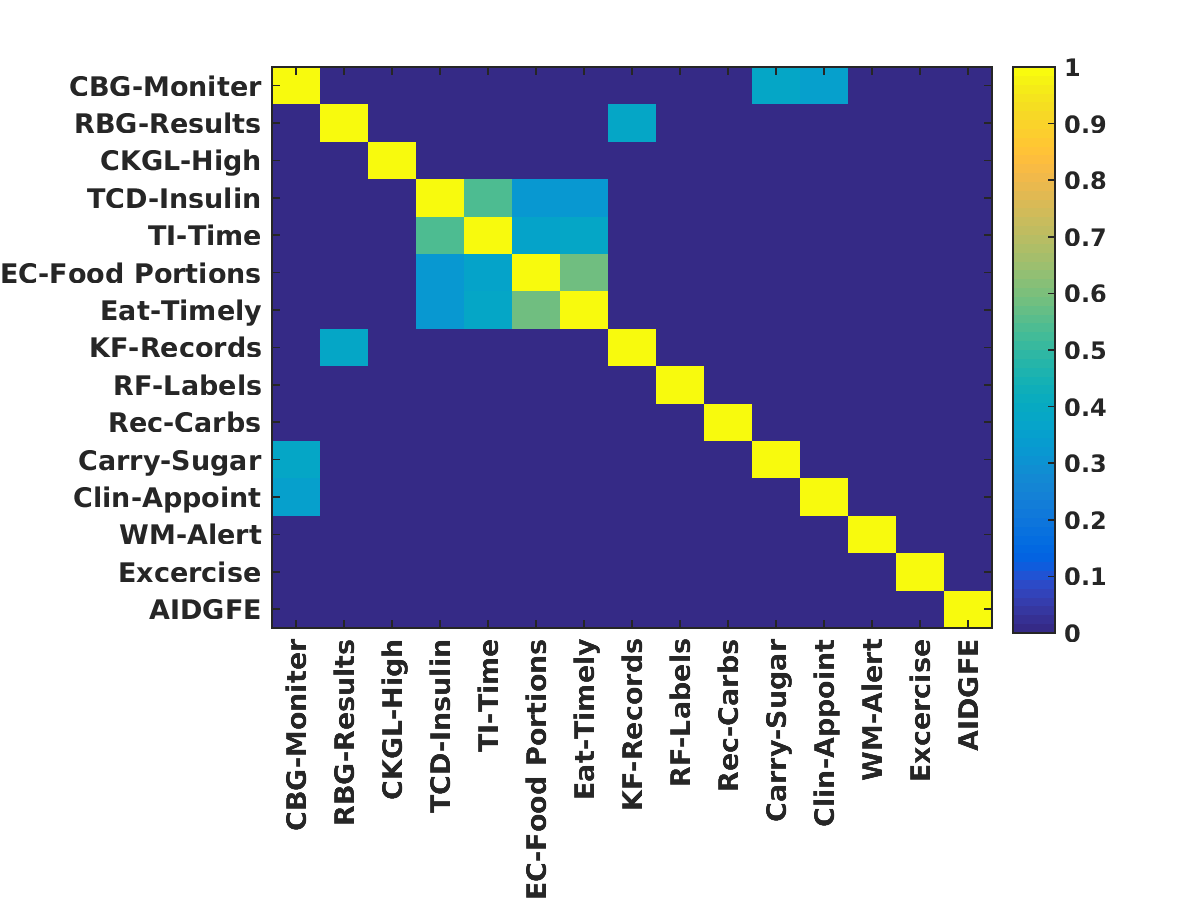}
\end{tabular}
\caption{Matrix of correlations greater than $0.3$ between the self-care factors.}
\label{fig_corr}
\end{figure}

\begin{figure}[h]
\centering
\begin{tabular}{c}
\includegraphics[width=3.5in]{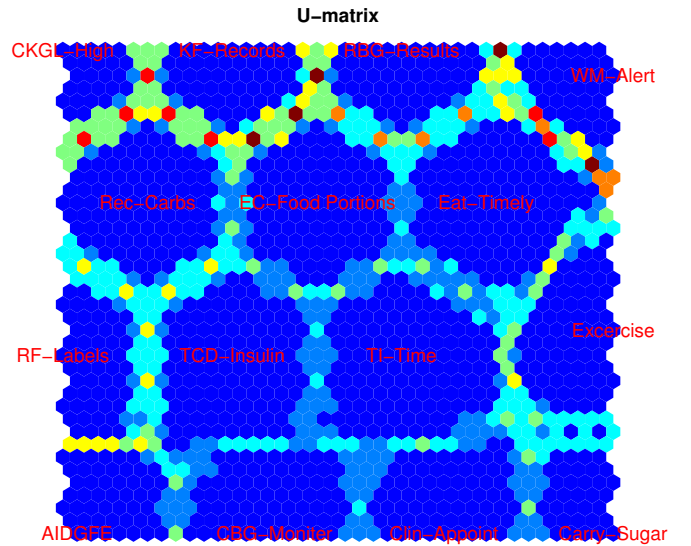}
\end{tabular}
\caption{U-matrix showing the self-care factor clustering produced by the SOM.}
\label{fig_umat}
\end{figure}

Respondent behaviours were analysed around the four areas covered by the survey: managing blood glucose, insulin intake, lifestyle and administrative. In order to investigate the relationships between self-care factors, the correlations between them were first computed on the data matrix (611$\times$15) of patient responses. The correlations found, after discarding weaker correlations (those less than $0.3$), can be seen in Figure~\ref{fig_corr}. In addition to analysing the correlations, the 611 patients' responses to the 15 survey questions were given as input to a SOM training algorithm\footnote{\url{http://www.cis.hut.fi/projects/somtoolbox/}}. The resultant U-matrix (Figure~\ref{fig_umat}) shows the results of clustering the self-care factors, with the distance between neurons indicated through their colour. Areas of dark blue indicate similar neurons, while lighter colours indicate reduced similarity and demarcate cluster boundaries. Representing the SOM via the U-matrix can therefore provide an intuitively appealing way to gain insight into the distribution of the self-care factors. In Figure~\ref{fig_umat}, the U-matrix is clearly divided into fifteen clusters, each of which represents a self-care factor. The clustering seen in the U-matrix shows strong agreement with the correlations seen in Figure~\ref{fig_corr}. For example, correlations can be seen between taking the correct dose of insulin (TCD-Insulin) and taking insulin at the correct time (TI-Time), eating the correct food portions (EC-Food Portions) and eating meals at the correct time (Eat-Timely), and between carrying quick acting sugars to treat low blood glucose (Carry-Sugar) and coming in for clinical appointments (Clin-Appoint). These correlations can be seen to hold in the U-matrix, with the clusters corresponding to each correlated pair being close together.

The output of the SOM can also be analysed using a component plane representation in order to analyse each self-care factor individually. A component plane view of the SOM can be thought of as a sliced version of the SOM with each plane containing the relative distribution of one self-care factor. By comparing component planes it is possible to see whether two components (self-care factors) correlate, with visually similar component planes indicating a correlation between their self-care factors. For example, in the components for taking the correct dose of insulin and taking insulin at the correct time are visually similar, and therefore correlate with one another (Figure~\ref{fig_ps_1}).

\subsection{Blood glucose}
\begin{figure*}[]
\setlength{\tabcolsep}{0.5pt}
\renewcommand{\arraystretch}{0.1}
\centering
\begin{tabular}{ccc}
\includegraphics[width=2.2in]{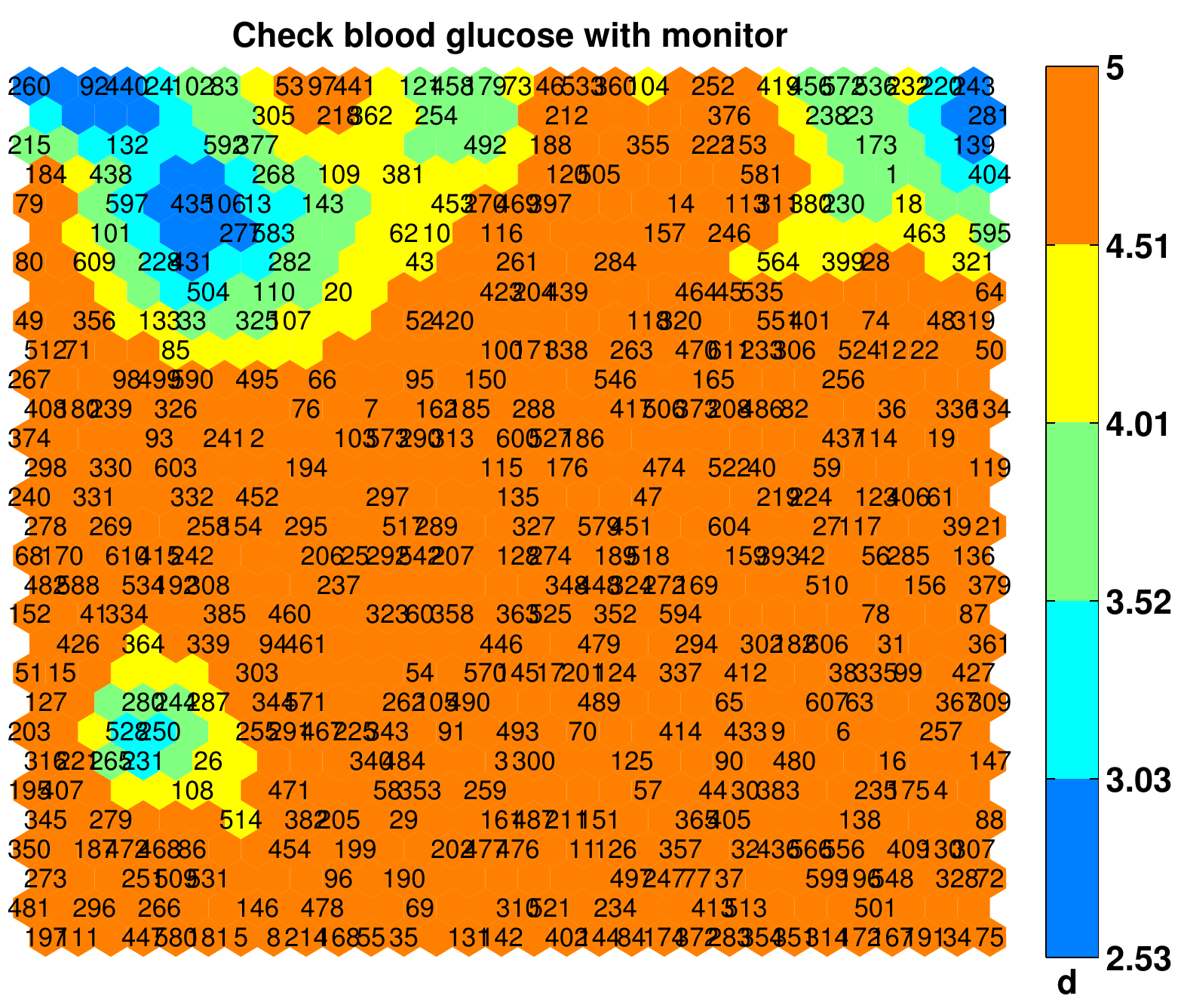} & \includegraphics[width=2.2in]{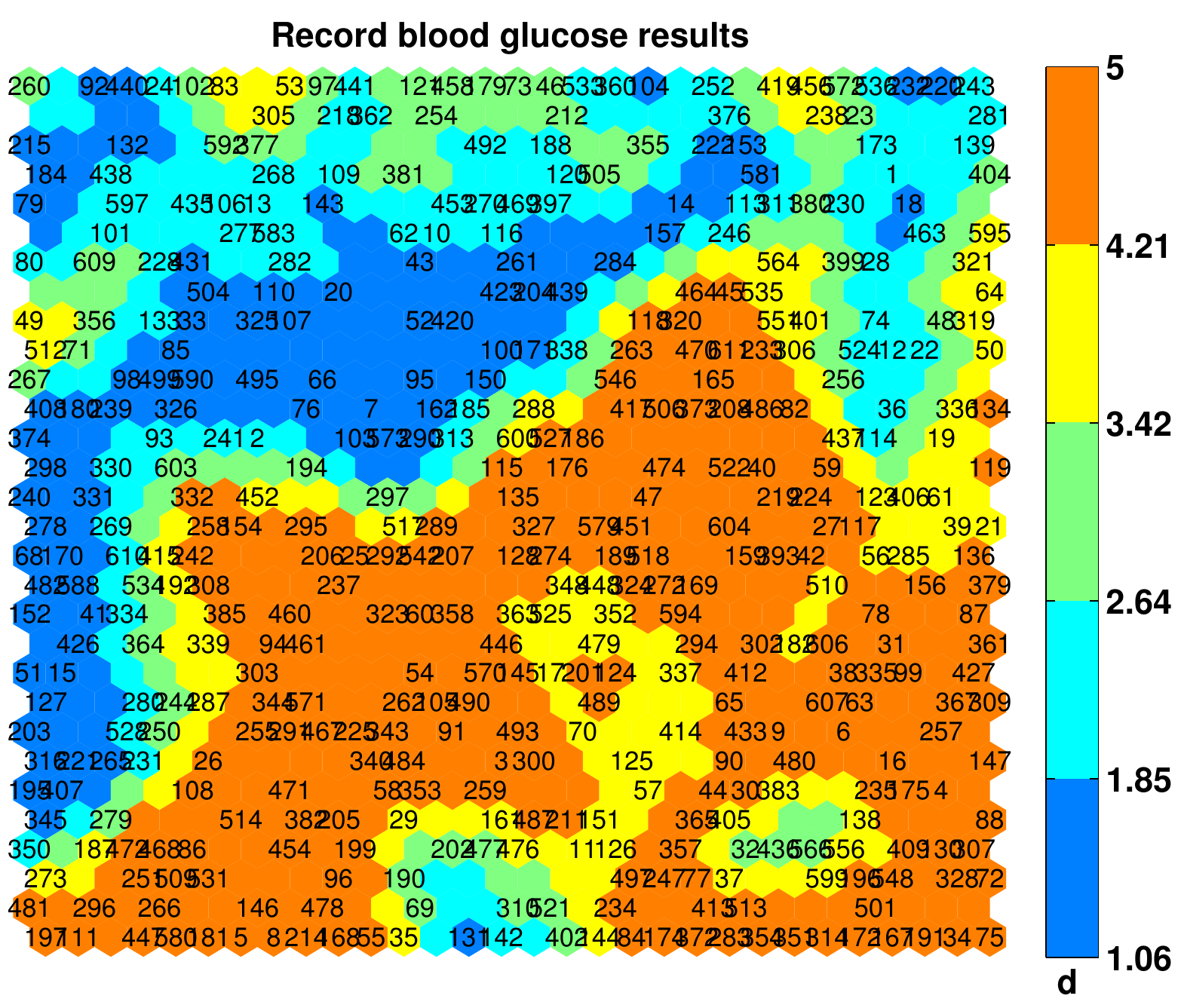} & \includegraphics[width=2.2in]{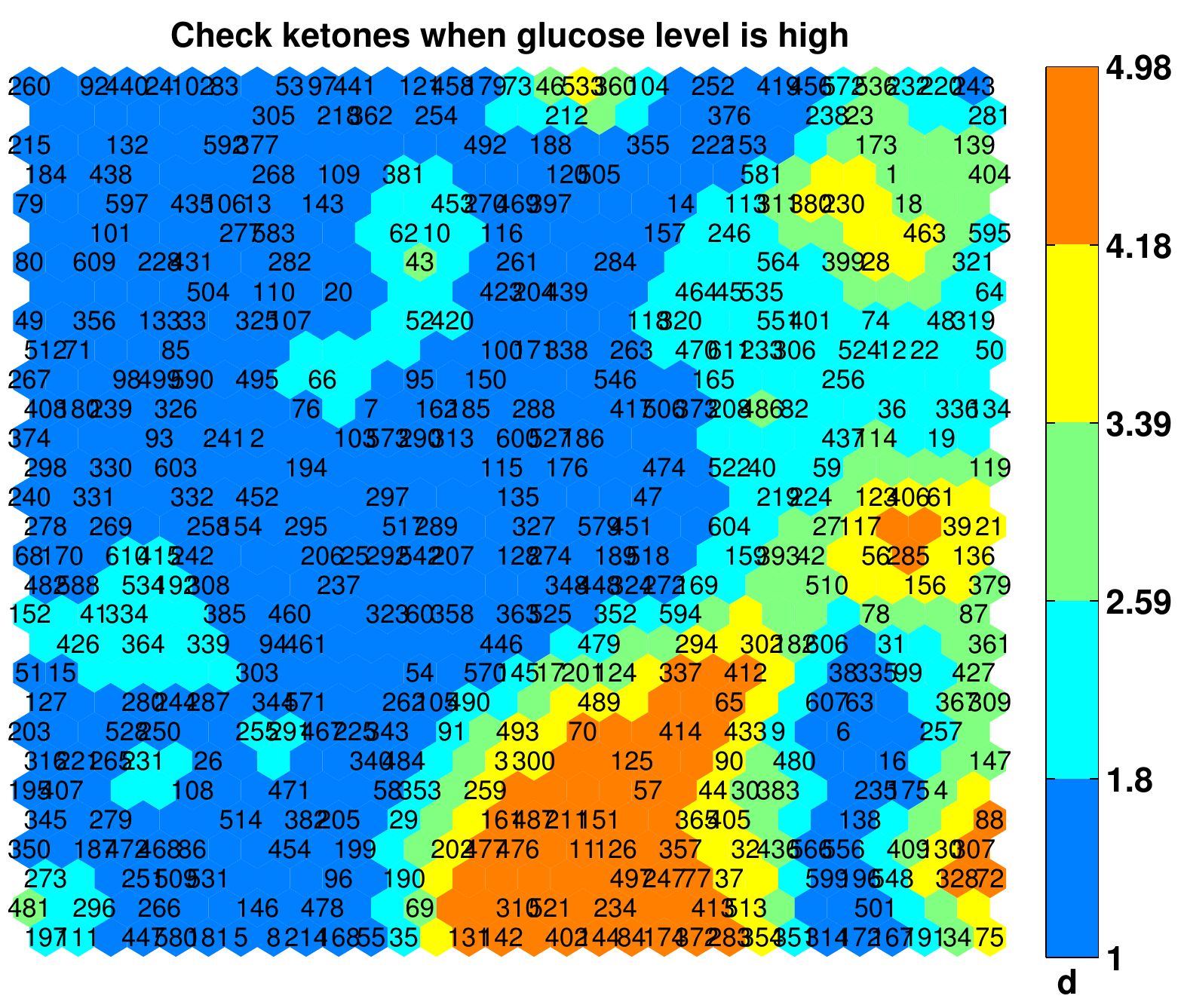}\\
\includegraphics[width=2.2in]{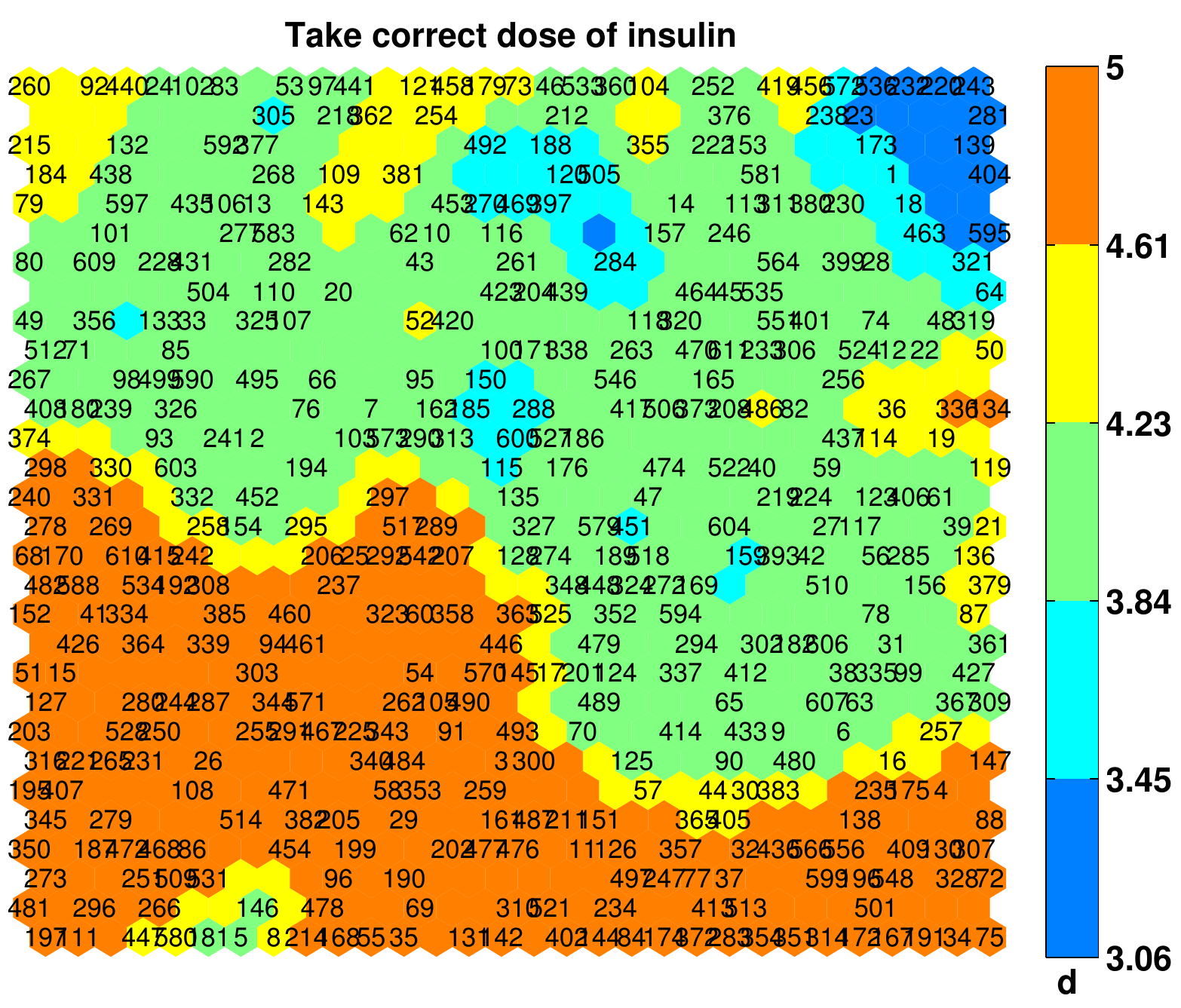} & \includegraphics[width=2.2in]{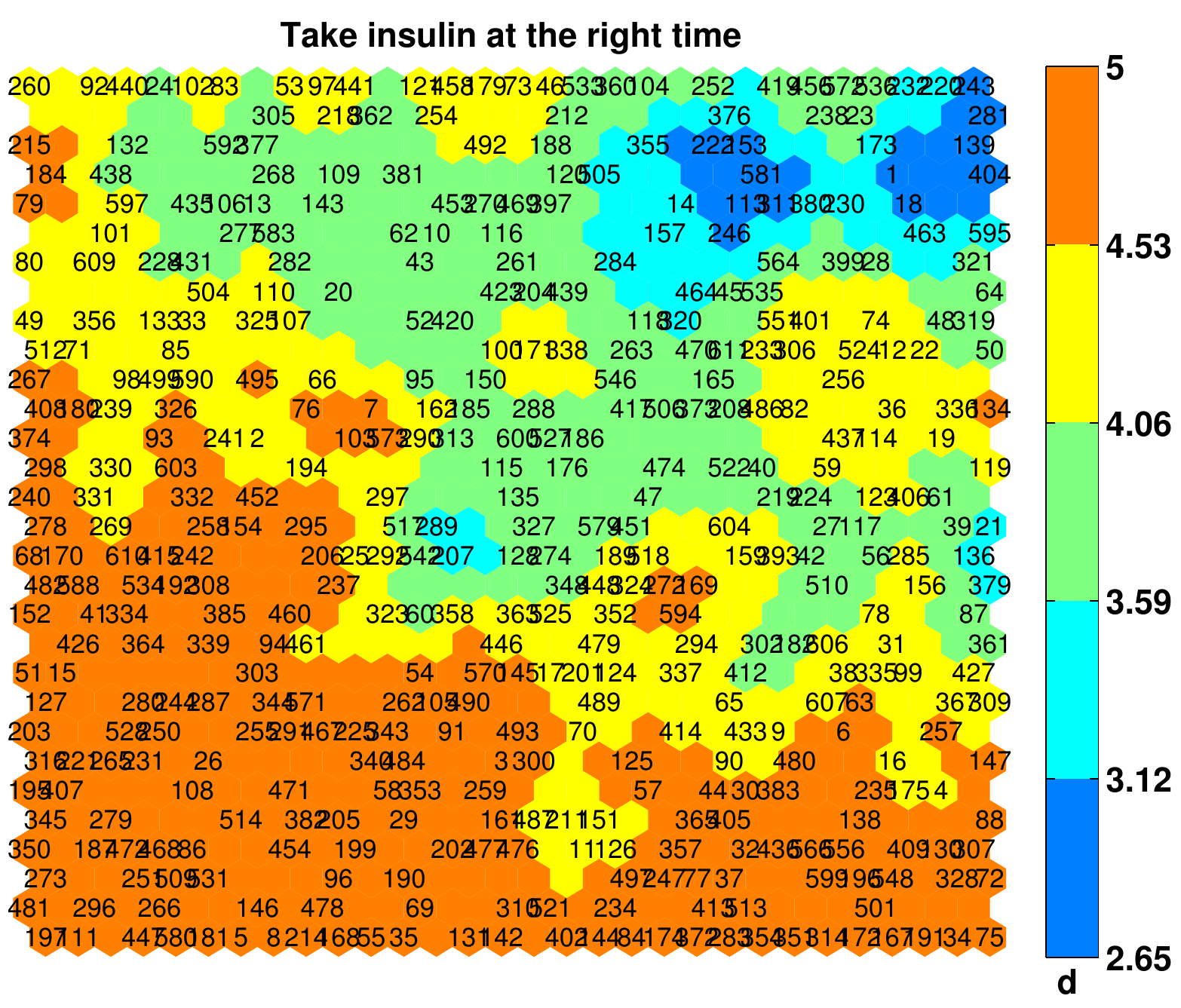} & \includegraphics[width=2.2in]{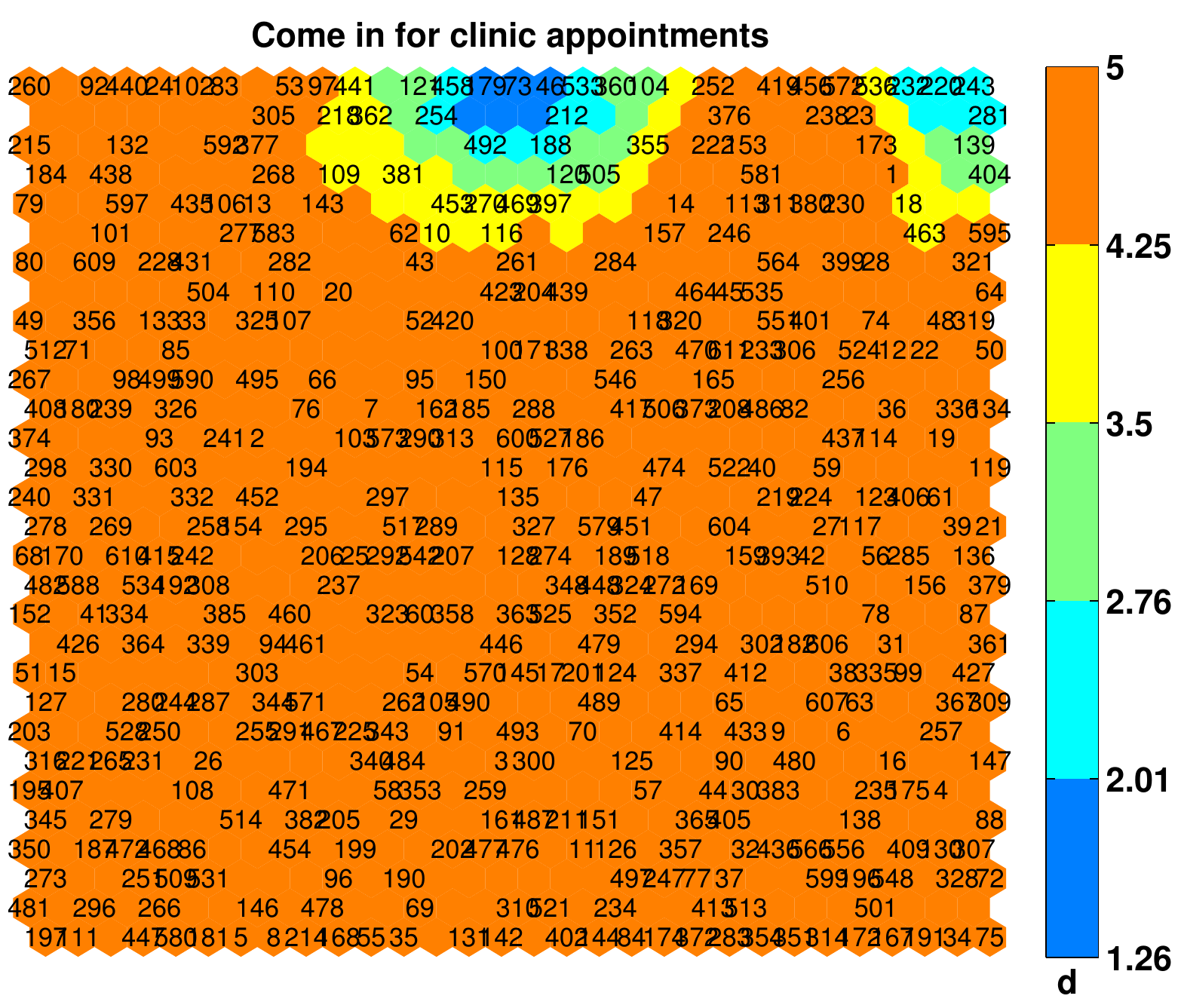}\\
\includegraphics[width=2.2in]{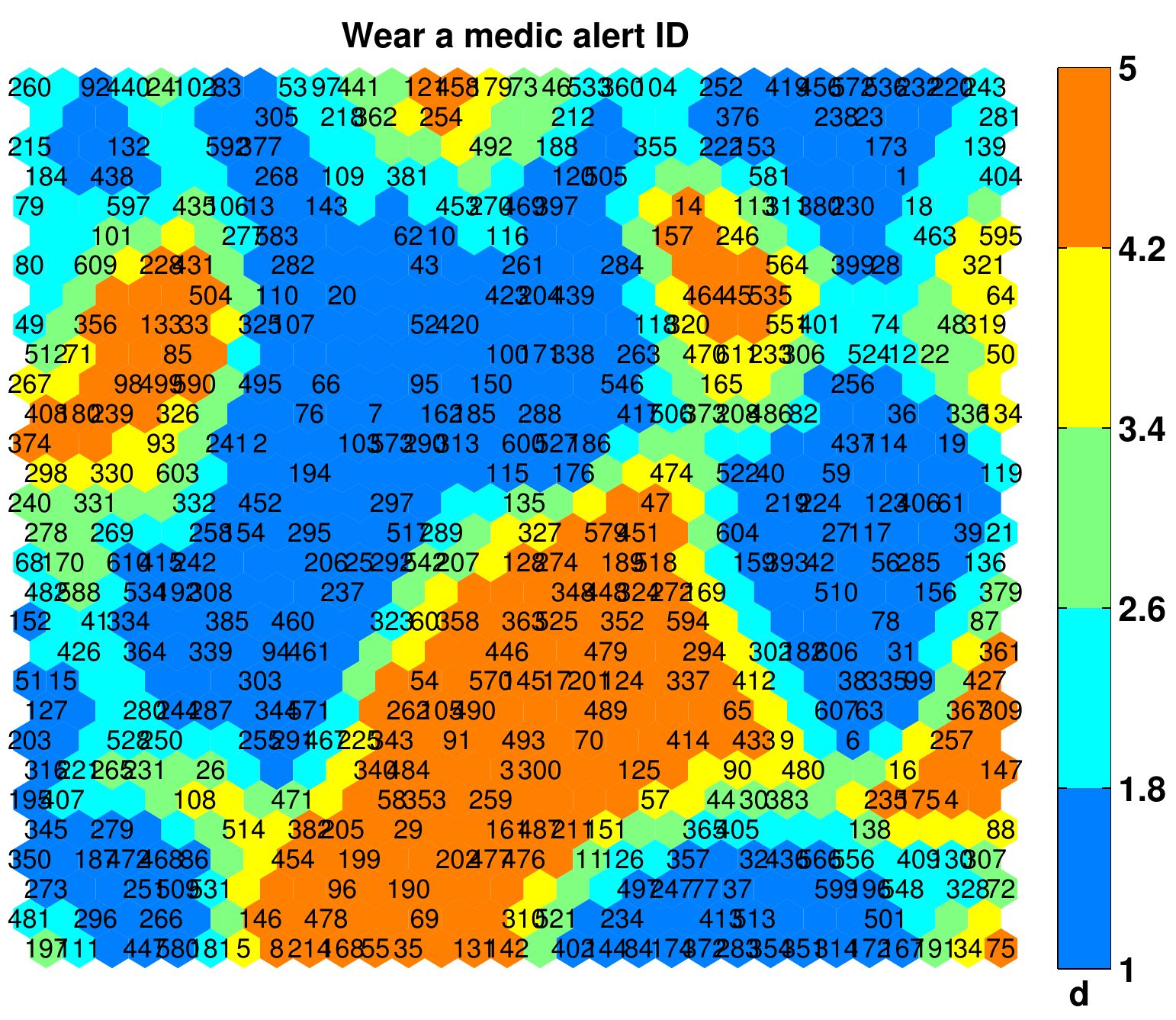} & \includegraphics[width=2.2in]{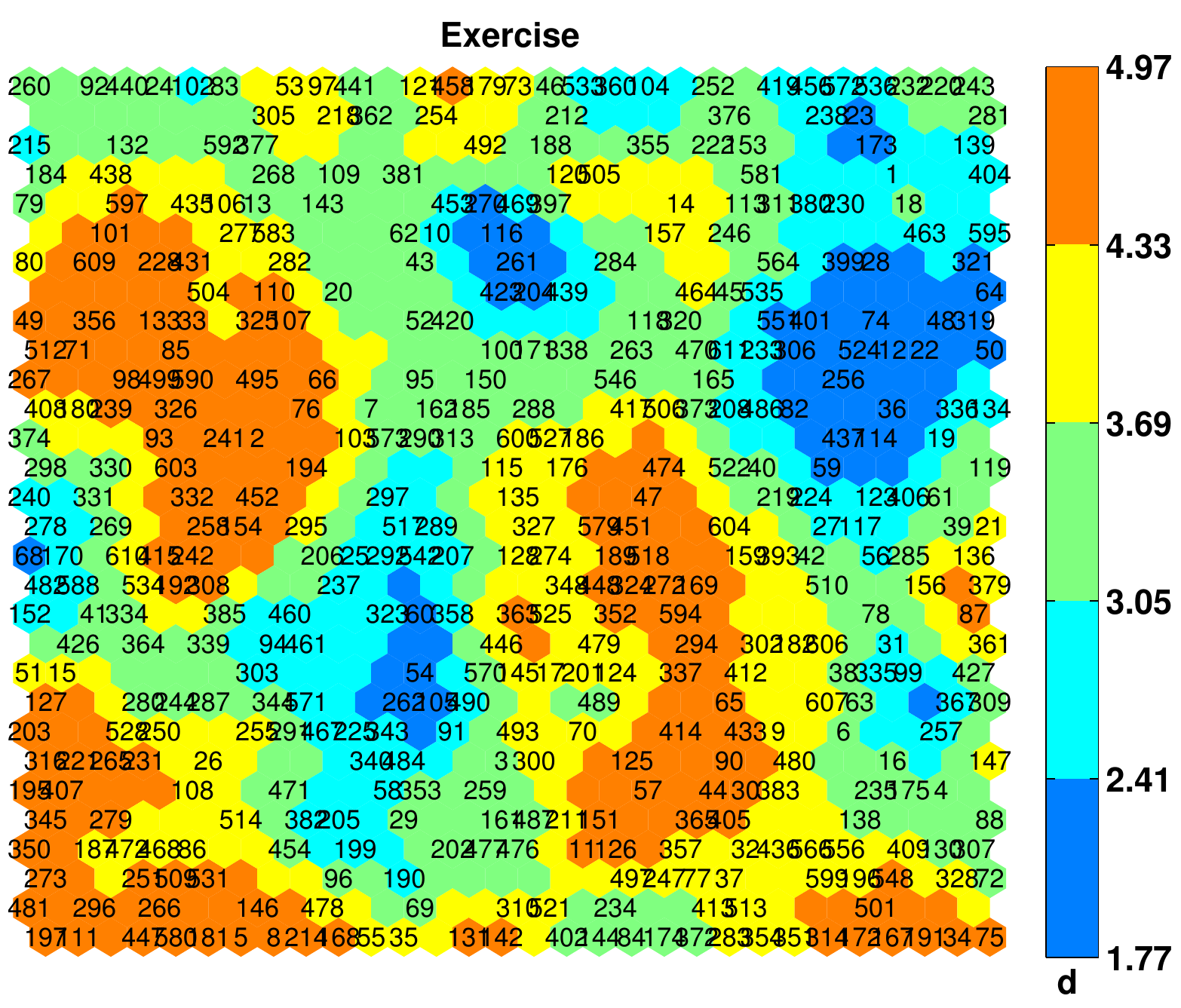} & \includegraphics[width=2.2in]{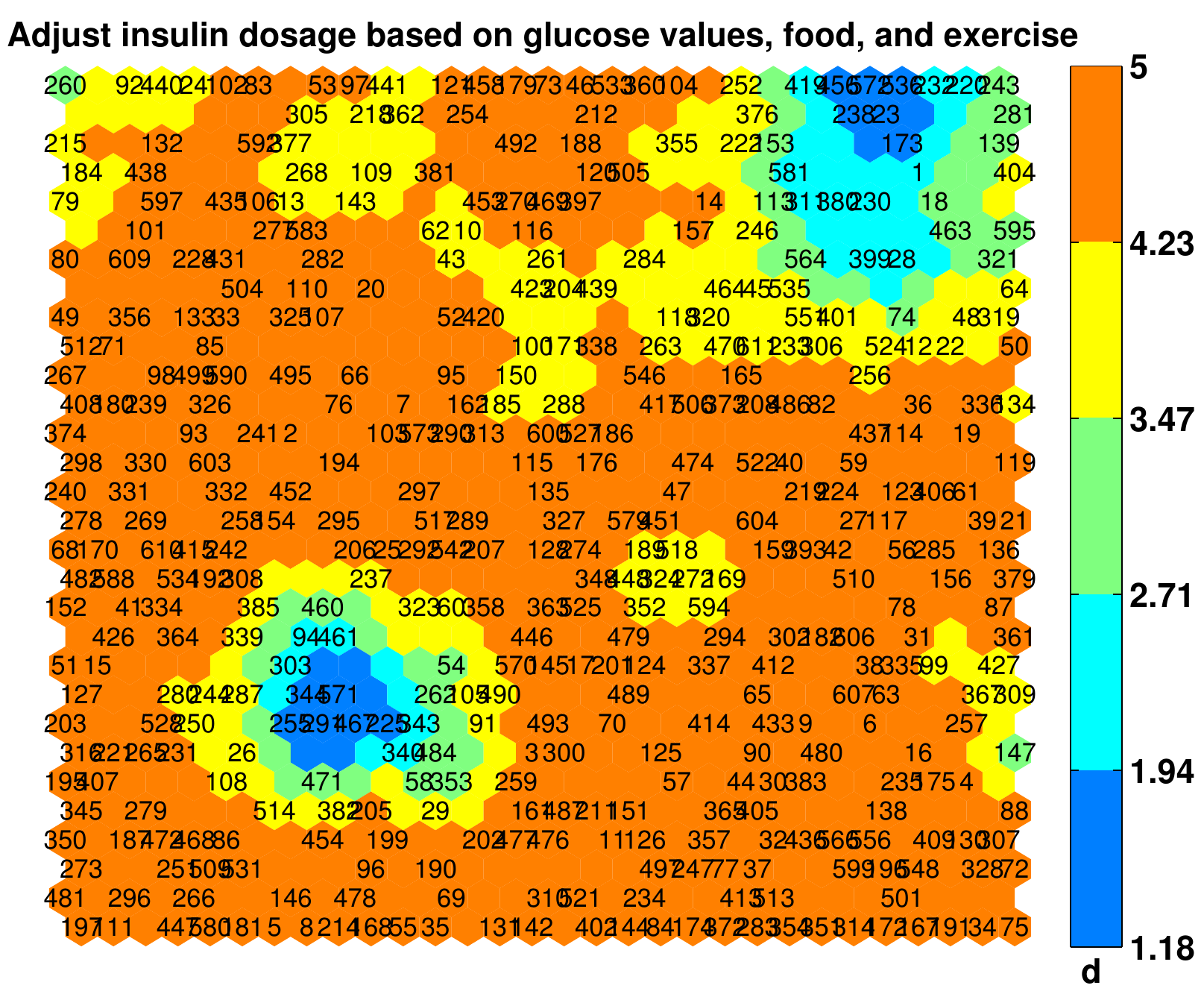}\\
\includegraphics[width=2.2in]{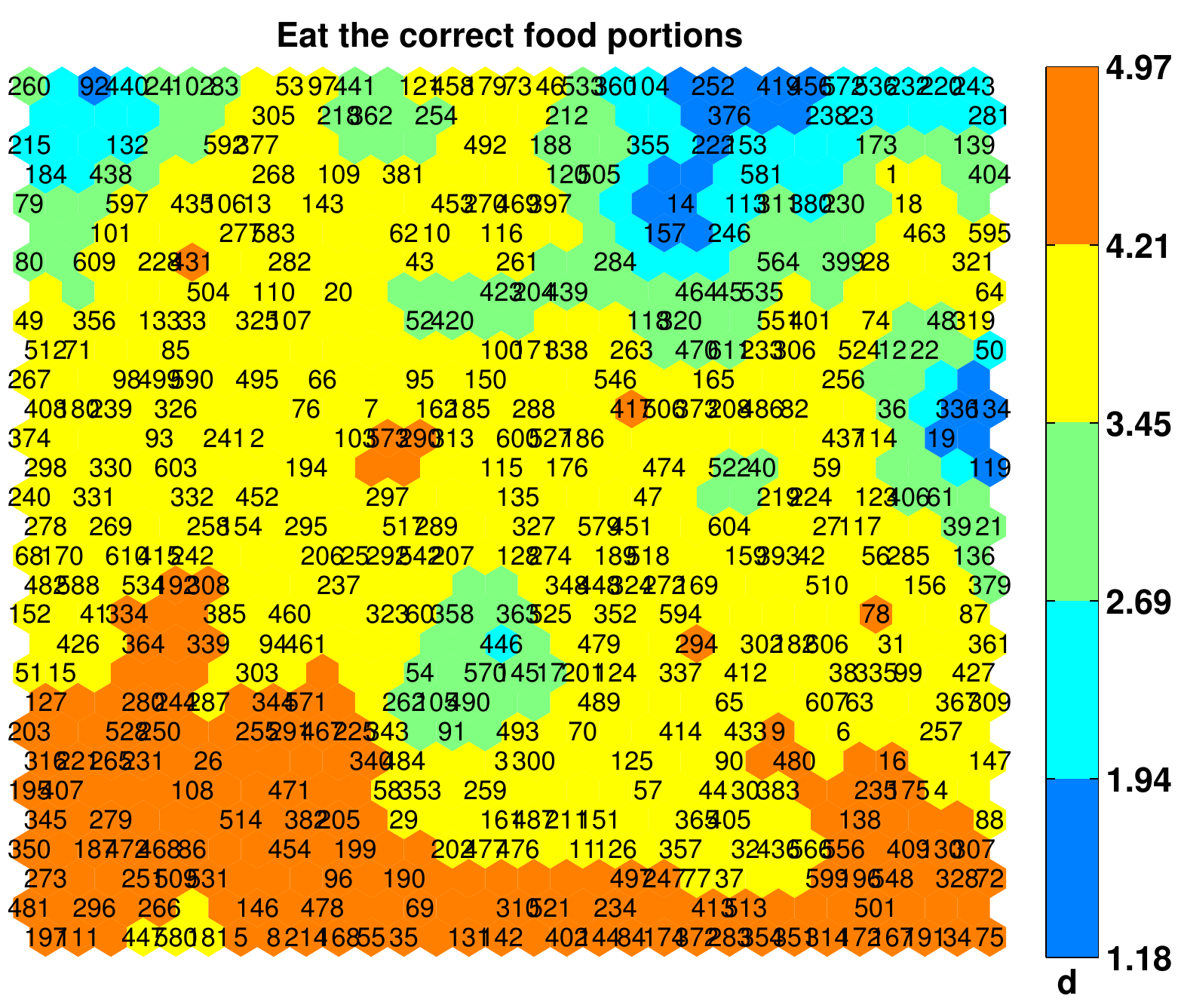} & \includegraphics[width=2.2in]{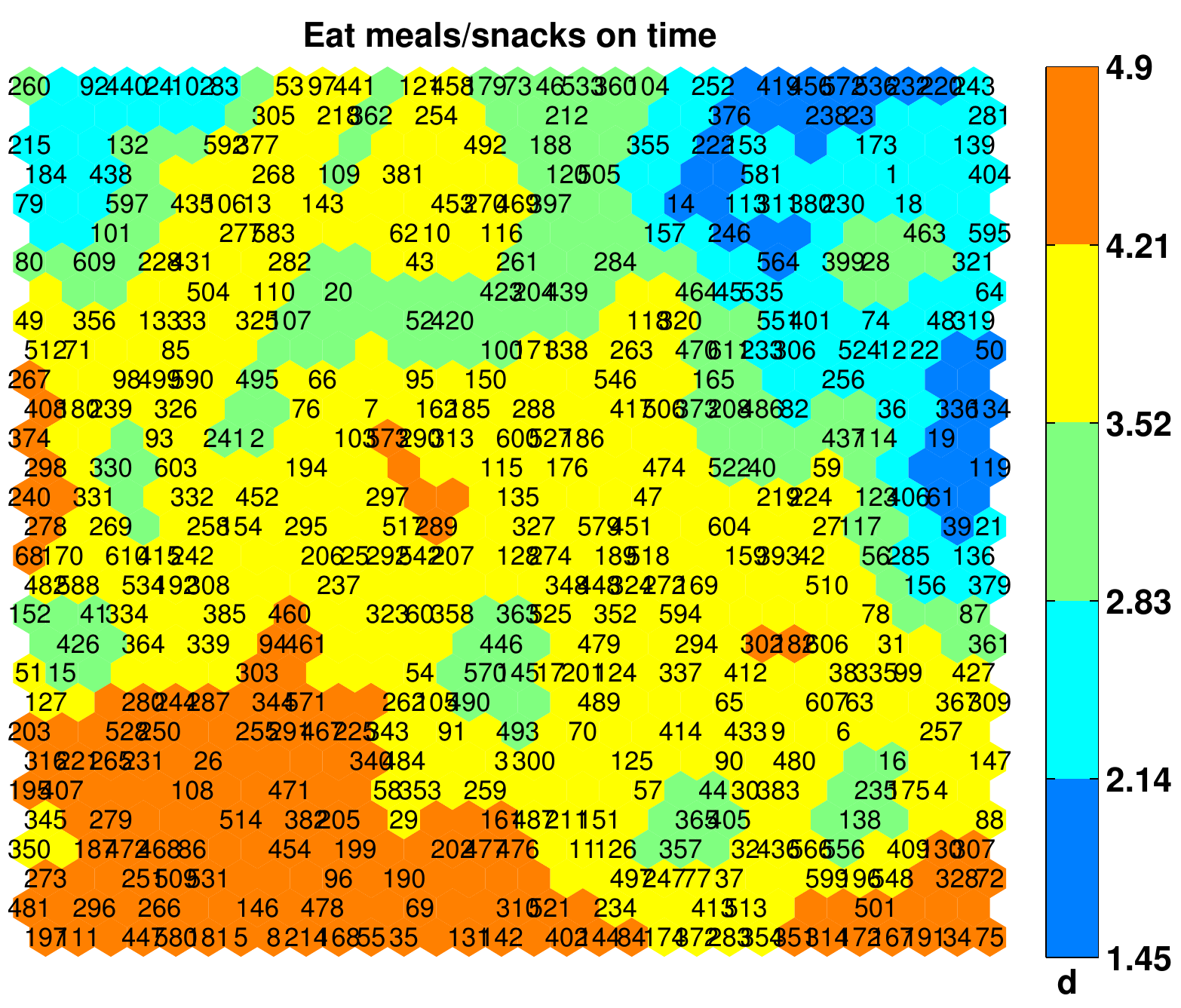} & \includegraphics[width=2.2in]{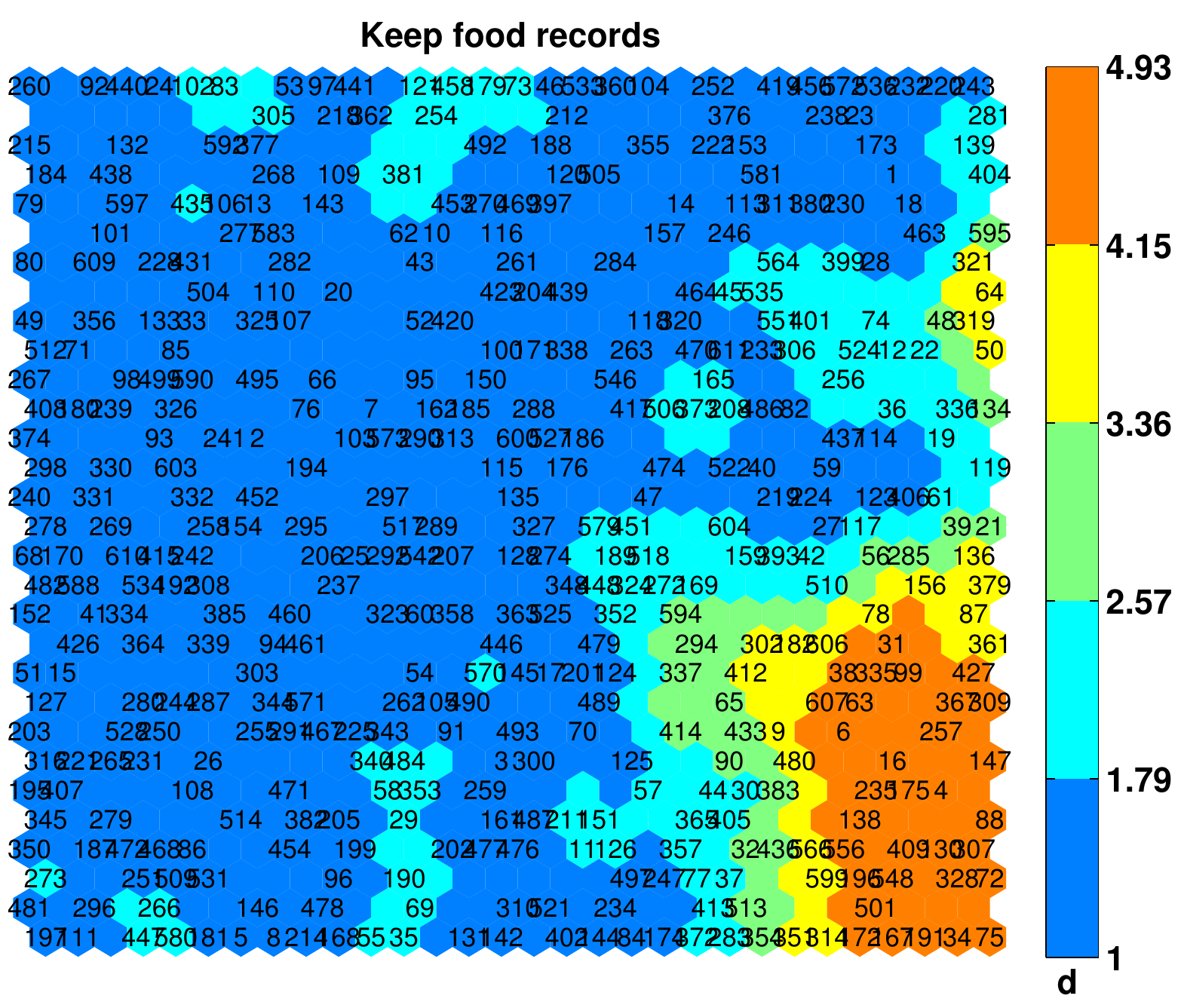}\\
\includegraphics[width=2.2in]{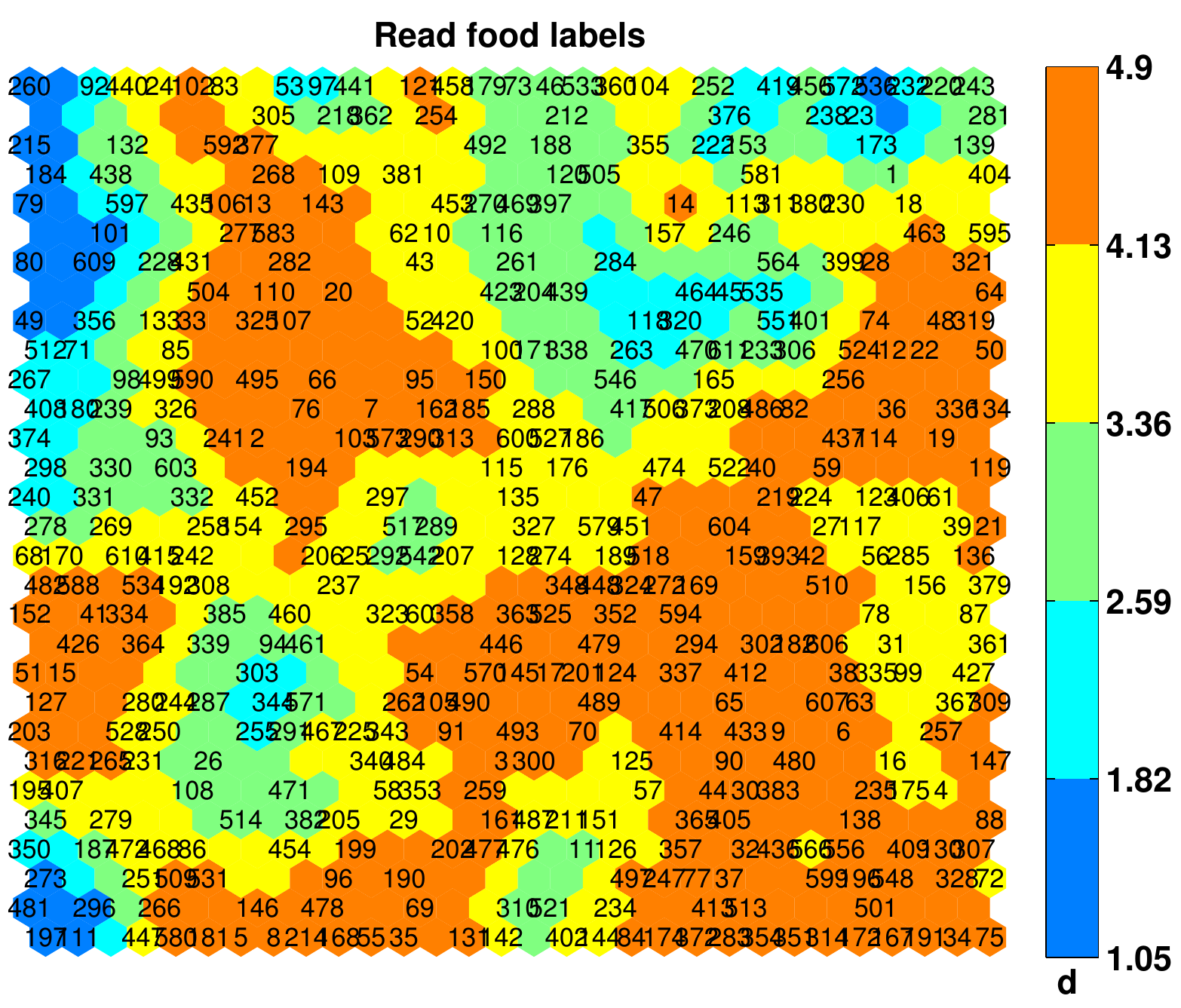} & \includegraphics[width=2.2in]{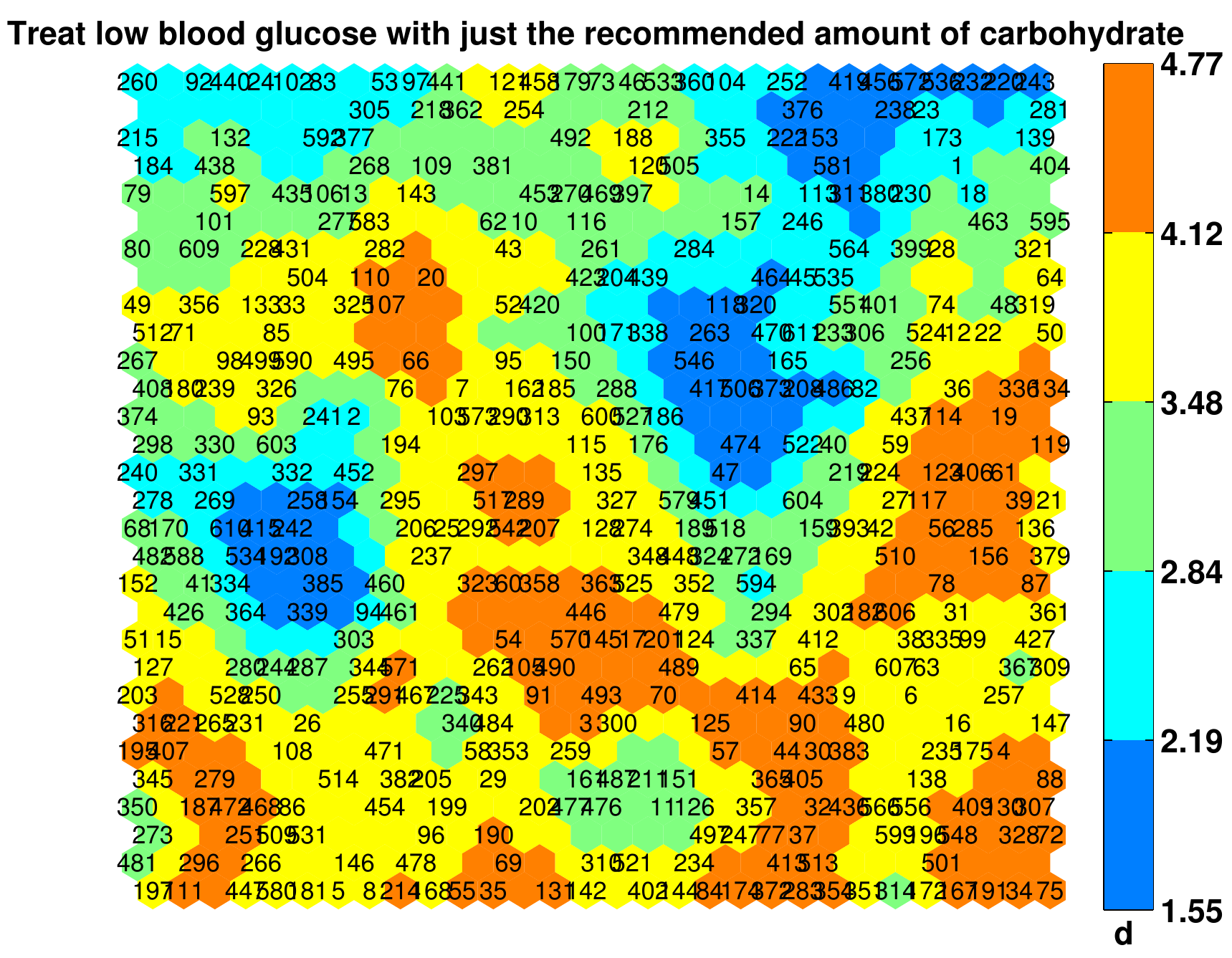} & \includegraphics[width=2.2in]{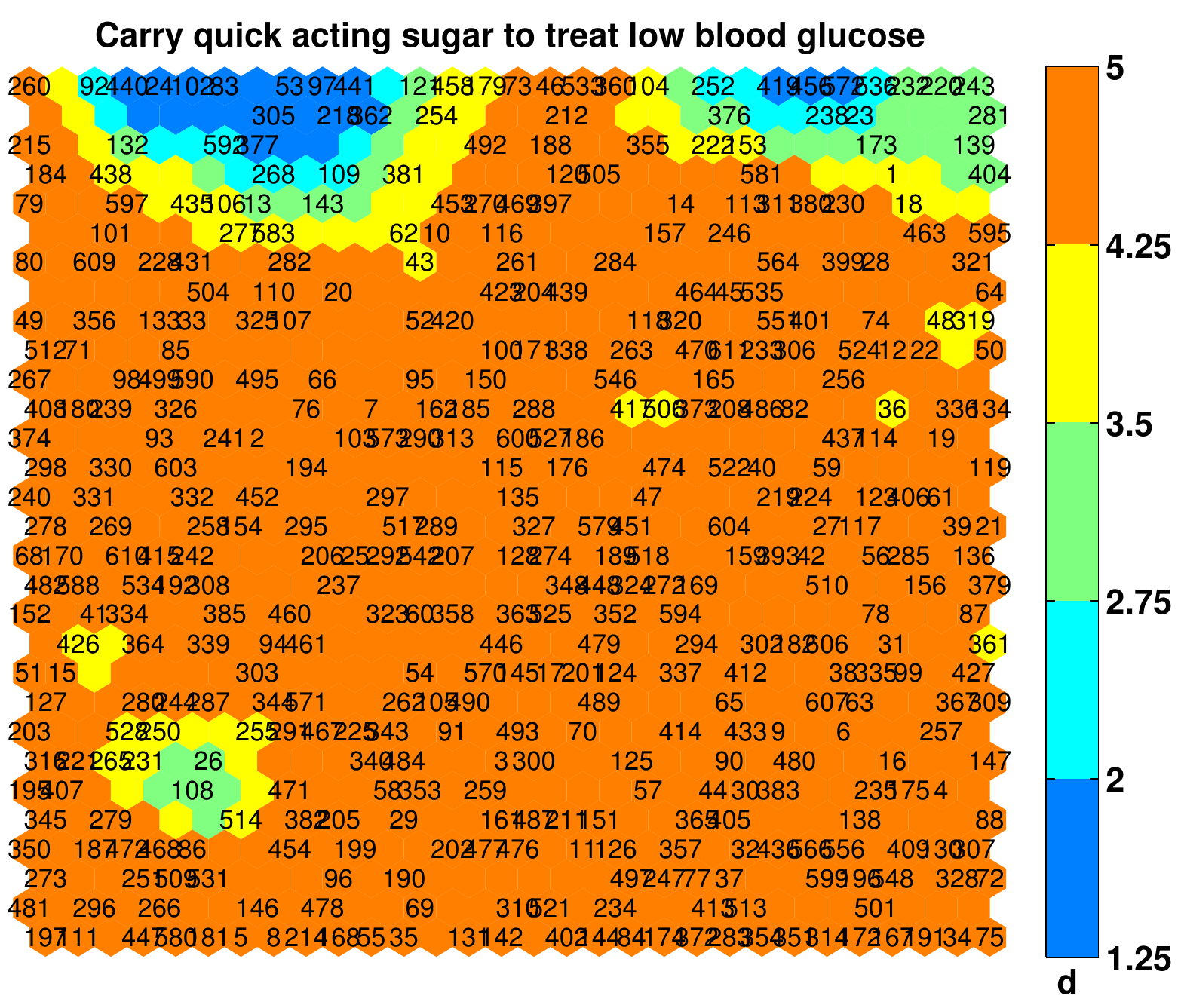}\\
\end{tabular}
\caption{Component plane representation of the SOM trained using patient survey responses. Patients are grouped according to the ratings given for each self-care factor, and each neuron is assigned a colour based on the grouped patients' responses, with a score of 5 (equivalent to a response of 'Always') corresponding to orange and the lowest score for the question corresponding to blue.}
\label{fig_ps_1}
\end{figure*}

From Figure~\ref{fig_ps_1}, we can see that the vast majority of respondents (more than 90\%) are usually or always checking their blood glucose levels, with over 50\% keeping a record of their glucose results. However, more than 90\% of respondents are not checking ketones when their blood glucose is high, potentially putting them at risk of diabetic ketoacidosis. There is therefore likely scope for further education of the risks posed by elevated blood sugar levels, especially with regards to the danger of elevated ketone levels.

\subsection{Insulin}
Approximately 50\% of patients can be seen to usually or always take the correct dose of insulin, with similar numbers taking insulin at the right time. Amongst those patients that do take the correct dose of insulin, the majority of them tend to also take their insulin at the correct time. Even though a sizeable fraction of respondents are only sometimes taking the correct dose of insulin, respondents were in general much better about adjusting their insulin dosages based on glucose levels, food and exercise.

\subsection{Lifestyle}
With the exception of exercise, patients tend to be conscientious when it comes to lifestyle-based management of their disease. The vast majority of patients carry quick acting snacks for periods of low blood sugar, and most patients usually or always read food labels, eat the correct portions of food and eat meals at the correct time. However, the more laborious diet-based self-care factors, treating with the right amount of carbohydrate and keeping food records, are poorly adhered to. In particular, keeping food records was the activity that the respondents did least frequently.

\subsection{Administrative}
Approximately 98\% of respondents always attend their clinical appointments, with similar response patterns to those patients that monitor their blood glucose and carry snacks to treat low blood sugar. However, very few people wear a medic alert ID more than sometimes.

\section{Conclusions and Discussions}
\label{condis}
This study demonstrates the capabilities of SOMs when attempting to infer useful information from survey data. In particular, when trying to understand the correlations between patient behaviours in an intuitive manner. For example, groupings of patients can be seen to be more or less conscientious in following their self-care treatments based on their answers to a few answers from the survey. Those patients who infrequently monitor their blood glucose, come in for clinic appointments and carry snacks to treat low blood glucose are also likely to respond 'Never' or 'Rarely' to the other questions. Clinicians can therefore potentially identify these patients as those in most need of greater intervention in their treatment, even in the absence of comprehensive answers to the other questions. Similarly, behavioural patterns leading to checking ketones, wearing a medic alert ID and keeping food records are poorly correlated with the others and extremely frequent. While these may be undesirable behaviours from a medical perspective, they do not provide much information as to a patients overall level of engagement with their self-care. Finally, although patients indicate that they are conscientious about adjusting their insulin levels, they are less good about taking the correct dose of insulin. Given the similarity between respondents who are likely to adjust their insulin dosage and those that monitor their blood sugar levels, the high rate of insulin dosage adjustment could be due to a greater awareness of their blood sugar levels on the part of the patients. However, given that reliability with which patients are taking the correct dose, it is possible that the adjustments made are not always correct.

Obtaining a better understanding of the patterns of behaviour that patients exhibit potentially presents clinicians with the opportunity to intervene in a patient's treatment, through lifestyle or medication adjustments, before they would otherwise have been able to identify the need. This is of particular benefit in situations where medical provision is scarce or costly, such as in less affluent areas and those without national health care~\cite{soltesz2009diabetes}. Therefore, tools that enhance clinicians' abilities to identify potentially detrimental behavioural patterns before they cause a deterioration in health can be used to guide early intervention plans, thereby reducing the burden on resources through a decrease in the need for costly or emergency treatment.


\section*{Acknowledgement}

The funding for this work has been provided by Department of Computer Science and Centre for Vision, Speech and Signal
Processing (CVSSP) - University of Surrey. `ST' and `SB' benefited from the Medical Research Council (MRC)-funded project `Modelling the Progression of Chronic Kidney Disease' (R/M023281/1): www.modellingCKD.org.



%
\bibliographystyle{IEEEtran}
\bibliography{santosh_references}

\end{document}